\documentclass[a4paper,11pt]{article}
\usepackage{aaskaiid}
\usepackage{array}
\usepackage{booktabs}
\usepackage{orcidlink}
\usepackage{rotating}
\usepackage{adjustbox}

\def\farcs{\hbox{$.\!\!^{\prime\prime}$}}

\title{Tracing the Origin of Circular-Symmetry Diffuse Radio Sources in the Next-Generation SKAO Era
}
\ShortTitle{Circular-Symmetry Diffuse Radio Sources}

\author[1]{Sabyasachi Pal\orcidlink{0000-0003-2325-8509}}
\ShortName{Pal et al.} 
\author[1]{Shobha Kumari\orcidlink{0000-0003-4213-9679}}
\author[1]{Souvik Manik\orcidlink{0000-0002-6794-7405}}
\author[2]{Marek Jamrozy\orcidlink{0000-0002-0870-7778}}
\author[3,4,5]{Ananda Hota\orcidlink{0000-0002-4959-7376}}
\affiliation[1]{Department of Pure and Applied Sciences, Midnapore City College, Kuturia, Bhadutala, Paschim Medinipur, West Bengal, 721129, India}
\emailAdd{sabya.pal@gmail.com}
\affiliation[2]{Obserwatorium Astronomiczne, Uniwersytet Jagielloński, ul. Orla 171, 30-244 Kraków, Poland}
\affiliation[3]{UM-DAE Centre for Excellence in Basic Sciences, University of Mumbai, Santacruz-East, Mumbai, 400098, India}
\affiliation[4]{Centre for Excellence in Theoretical and Computational Science, University of Mumbai, Santacruz-East, Mumbai, 400098, India}
\affiliation[5]{RAD@home Astronomy Collaboratory, Kharghar, Navi Mumbai, 410210, India}
\abstract{Circularly symmetric diffuse radio sources represent a recently recognized and rare class of extragalactic synchrotron-emitting structures whose physical origin remains uncertain. The currently known population is small, and the limited frequency coverage, sensitivity, and polarimetric information available from existing observations hinder robust discrimination among the various proposed formation scenarios. This chapter presents an overview of four representative circular diffuse sources featuring diverse morphologies, such as a horseshoe-shaped inner ring, circular diffuse emission surrounded by inner S/Z-shaped structures and a double-ring diffuse source. The angular sizes of these sources are in the range of 60--100 arcsec, similar to odd radio circles (ORCs). These systems illustrate the morphological diversity of circular diffuse radio emission and provide valuable clues to the physical processes responsible for their formation. The Square Kilometre Array Observatory (SKAO), with its unprecedented sensitivity, wide instantaneous bandwidth, sub-arcsecond imaging capability, and high-fidelity polarimetric performance, will significantly advance the study of these objects. SKAO surveys are expected to detect a substantially larger population of faint circular diffuse radio sources, enabling the first statistically meaningful studies of their occurrence, environments, and evolution. An expanded sample will allow investigations of their cosmological evolution and help determine whether their origin is linked to AGN feedback processes, episodic jet activity, galaxy interactions, or large-scale structure formation. This chapter summarizes the current observational knowledge of circular diffuse radio sources and describes how next-generation radio facilities, particularly the SKAO, will enable progress in understanding the origin and physical nature of these enigmatic radio structures.}

\begin{document}
\maketitle

\section{Introduction}
Radio galaxies powered by active galactic nuclei (AGN) exhibit a wide variety of morphological structures. Most of these systems contain a compact radio core that coincides with the optical nucleus of the host galaxy. The overall morphology of a radio galaxy is shaped primarily by the interaction of relativistic jets with the surrounding intergalactic and intracluster medium, as well as with the hot diffuse gas that envelops the host. These jets consist of highly energetic magnetized plasma, and they can propagate from parsec-scale regions near the central engine out to hundreds of kiloparsec (kpc) into the intergalactic medium (IGM).

When the jet encounters the surrounding medium, it undergoes recollimation, leading to the formation of internal shocks that decelerate the jet and promote mixing with the ambient material \citep{Pe07}. At the centre of these active systems lie supermassive black holes (SMBHs) with masses in the range of $10^6$$-10^{9.5}$ M$_{\odot}$, typically hosted by the bulges of massive elliptical galaxies. The feedback processes driven by accretion onto these SMBHs play a pivotal role in regulating both the morphology and evolution of radio sources and in influencing the co-evolution of galaxies and their environments \citep{Ko95, Si98, Ki03, Gr04, Sp05, Ho06}.

Recent wide-field radio surveys have revealed a variety of diffuse radio sources with unusual morphologies that do not fit easily within the known classification of radio galaxies. Among these are large-scale circular radio structures with diameters of several hundred kpc, often surrounding a central galaxy but lacking the prominent jet–lobe morphology typically observed in classical radio galaxies. Such structures suggest that energetic processes in galaxies and their environments may produce quasi-spherical distributions of synchrotron-emitting plasma.

One notable class of such objects is the odd radio circles (ORCs) \citep{No21b, No21c, Ko21, Om22c, Lo23, koribalski2024, Ko24a, No25}, which appear as ring-like or edge-brightened radio features on scales of 200--500 kpc. Several candidate circular radio sources have recently been identified in large-area surveys such as the Evolutionary Map of the Universe (EMU), where automated detection techniques reveal diffuse ring-like features with low surface brightness \citep{Gu25}. 
A particularly intriguing subset of circular radio sources exhibits double ring structures \citep{2022MNRAS.513.1300N, Ta25, 2025MNRAS5431048H}. Such morphologies may indicate multiple episodes of energetic activity or sequential shock fronts propagating through the surrounding medium.  Although their detailed origin remains under active case of study, ORCs are thought to be related to energetic processes associated with galaxy evolution and AGN activity. Proposed scenarios include expanding shock fronts generated by AGN outflows, relic radio plasma re-energized by compression or turbulence, and large-scale interactions within galaxy groups or clusters.

Circular diffuse radio sources, therefore, provide a valuable opportunity to probe large-scale energy release processes associated with AGN activity, feedback, and environmental interactions. Understanding the origin and evolution of these structures can offer important insights into the interplay between galaxies, relativistic plasma, and the surrounding intergalactic medium.

\begin{figure*}
    \vbox{
    \centerline{
    \includegraphics[width=7.8cm]{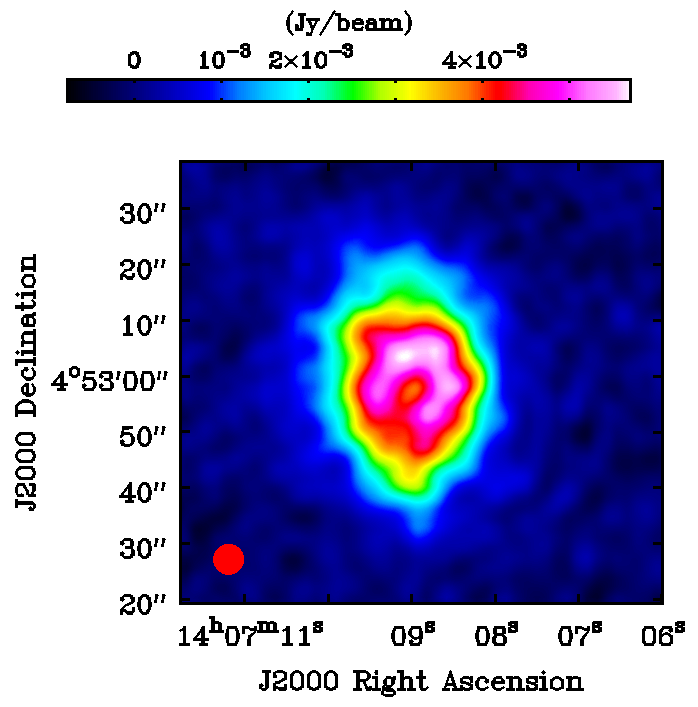}
    \includegraphics[width=7.5cm]{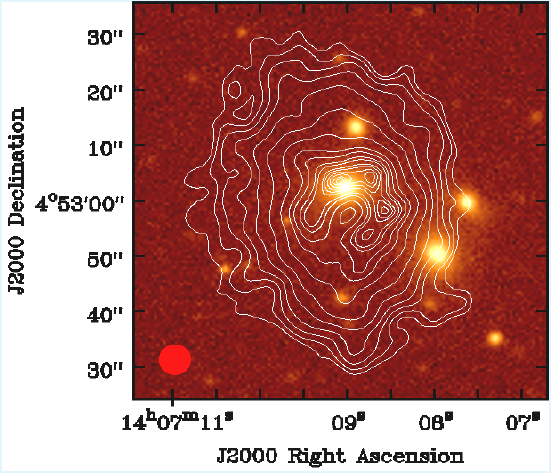}
    }
    }
    \caption{Left: VLA FIRST radio image of the circular diffuse radio source J1407+0453, showing diffuse circular emission with a bright central component. Right: Radio–optical overlay of the source, where VLA FIRST radio contours are superimposed on the Pan-STARRS1 optical r-band image. The contours reveal a compact central galaxy embedded within diffuse circular radio emission and a horseshoe-shaped inner ring surrounded by extended radio structure. These images are adapted from \citet{Ku23b}.}
    \label{fig:J1407+0453}
\end{figure*}

\section{Recent discovery of circular diffuse radio sources}

In this section, we discuss three circular diffuse sources \citep{Ku23a, Ku23b, Ku25} identified using the Very Large Array (VLA) Faint Images of the Radio Sky at Twenty cm \citep[FIRST;][]{Be95, Wh97} survey and the LOw-Frequency ARray \citep[LOFAR;][]{vanHaarlem2013} Two-metre Sky Survey second data release \citep[LoTSS DR2;][]{Shimwell2022}. 

\subsection{J1407+0453: Horseshoe-shaped ring of diffuse emission}
\citet{Ku23b} recently reported a remarkable source exhibiting a horseshoe-shaped ring (HSR) surrounded by extended diffuse radio emission (see Figure~\ref{fig:J1407+0453}). The angular size of the system is about 65 arcsec ($\sim 160$ kpc in projection at $z = 0.13360$), whereas the diameter of the HSR is approximately 10 arcsec (25 kpc in projection). Although the overall circular appearance places it close to the known population of ORCs \citep{No21b, No21c, Ko21, Om22c, Lo23, koribalski2024, Ko24a, No25}, the presence of a distinct inner ring embedded within diffuse emission makes this source particularly unusual. To date, such an inner ring morphology has not been commonly observed among previously reported circular sources.

The bending in HSR of J1407+0453 looks similar to the bending we typically observed in narrow-angle tailed \citep[NAT;][]{Bliton1998, Patra2019} radio galaxies. NAT sources are generally found in dense cluster environments where ram pressure, generated by the motion of the host galaxy through the intracluster medium, bends the radio jets into narrow tails (viz. \citet{Bhukta2022, Sasmal2022, Pal2023}). However, J1407+0453 does not appear to be associated with a cluster environment, making the origin of the observed bending less straightforward. The presence of a compact inner ring embedded within the diffuse structure suggests that multiple dynamical processes may be shaping the radio morphology. High-resolution and multi-frequency radio observations will be important to better characterise the spectral properties and internal structure of this object and to determine whether J1407+0453 represents a distinct morphological subclass among circular diffuse radio sources.

\subsection{Circular diffuse sources with internal radio structure}

While some circular radio sources exhibit nearly symmetric ring-like morphologies, others display diffuse or irregular structures (see Figure \ref{fig:VLA_C_J1218+1813}) that deviate from an ideal circular shape \citep{Ku23a, Ku25, Gu25}. These systems often show variations in surface brightness, fragmented edges, or partially filled interiors, indicating a complex internal structure within the diffuse emission.

The circular diffuse sources J1507+3013 and J1218+1813 reported in \citet{Ku23a} and \citet{Ku25} have angular sizes comparable to those of ORCs \citep{No21b, No21c, Ko21, Om22c, Lo23, koribalski2024, Ko24a, No25}, but display markedly different internal radio structures. In contrast to the edge-brightened rings typically associated with ORCs, these sources exhibit significant radio emission within the interior of the circular structure.

In particular, both systems contain radio features resembling lobe-like or S-shaped structures embedded within the surrounding diffuse emission. The internal morphology observed in J1507+3013 and J1218+1813 is reminiscent of the structures seen in winged radio galaxies. High-resolution imaging of J1218+1813 (see Figure~\ref{fig:VLA_C_J1218+1813}) reveals a complex radio morphology that bears similarities to S/Z-shaped radio galaxies reported in previous studies \citep[e.g.][]{Misra2023}.

Similar asymmetric circular diffuse sources have also been identified in recent large-area survey catalogues such as that presented by \citet{Gu25}. These findings suggest that circular diffuse radio emission can be associated with a variety of internal radio structures and that the morphological diversity among such sources is significantly larger than initially recognized.

\subsection{RAD@home discovery of extragalactic double radio rings}
\begin{figure}[h]
\vbox{
    \centering
	\includegraphics[width=10cm]{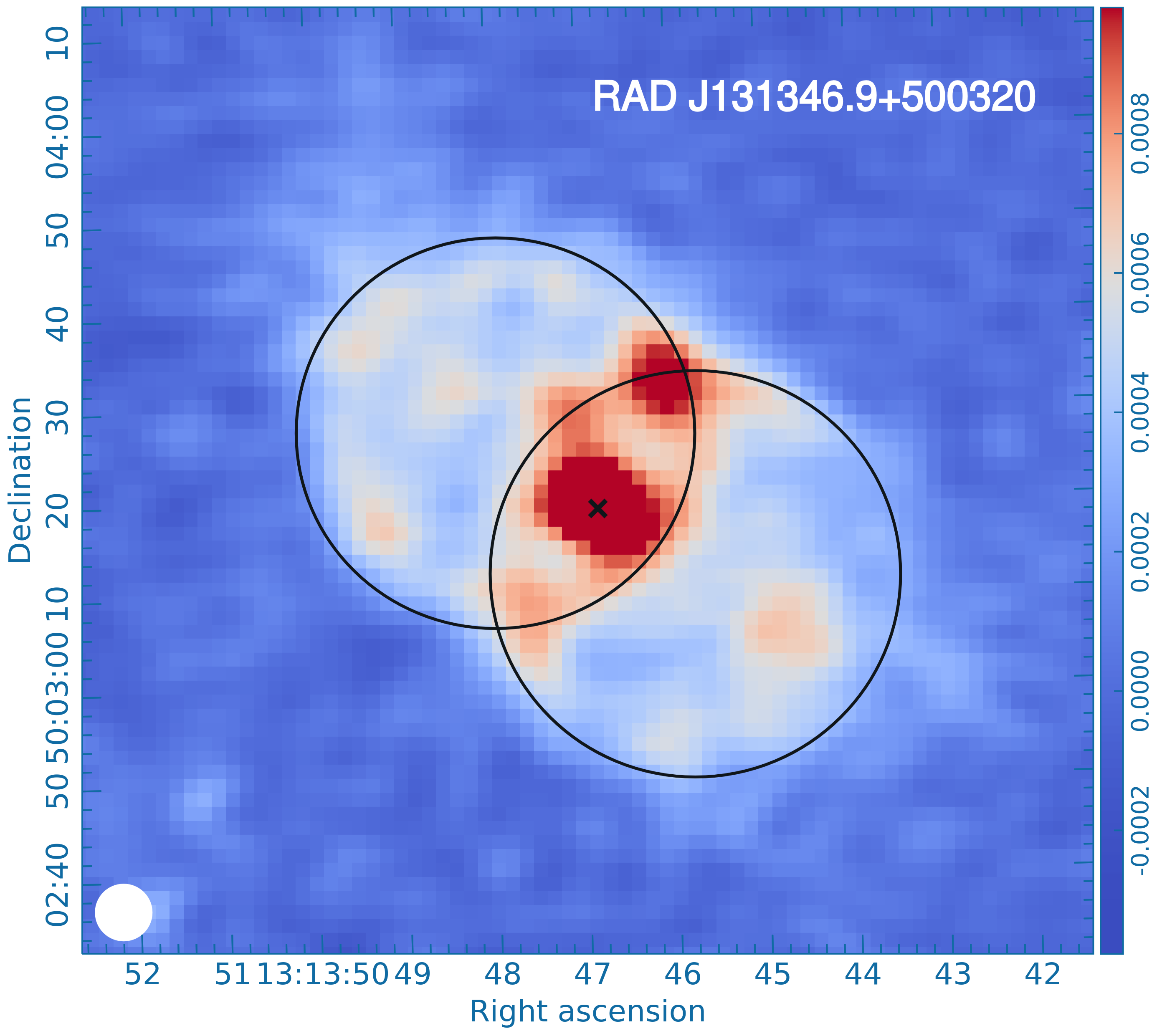}
    }
    \caption{Example of a double-ring circular radio structure, RAD J131346.9+500320, showing two partially overlapping diffuse rings centered on a host galaxy (marked by the cross), adapted from \citet{2025MNRAS5431048H}.}
    \label{fig:double_ring_orcs}
\end{figure}

A particularly intriguing example of circular radio morphology is the recently discovered double-ring system RAD J131346.9+500320 reported by \citet{2025MNRAS5431048H} (see Figure~\ref{fig:double_ring_orcs}) from the LoTSS DR2. The radio image reveals two partially overlapping circular structures with a total angular extent of $\sim 100$ arcsec, corresponding to 800 kpc in projection (measured from the LoTSS 6 arcsec). The morphology of RAD J131346.9+500320 is characterized by two diffuse ring-like structures with enhanced radio brightness along their rims. Edge-brightened rims are often interpreted as signatures of compression or shock fronts, where relativistic electrons are re-accelerated, and magnetic fields are amplified. The partial overlap of the rings may indicate that the expanding structures are not perfectly concentric but instead offset along a preferred axis relative to the host galaxy. 

A possible explanation for the formation of double-ring radio structures involves the action of a bipolar superwind originating from the host spiral galaxy after the original radio lobes have evolved into a relic or remnant phase. In this situation, large-scale outflows driven by the superwind could interact with the previously existing relic radio plasma, producing shock fronts that manifest as two expanding circular radio rings. While single-ring ORCs may arise from the re-energization of aged radio lobes associated with elliptical host galaxies, the generation of twin-ring systems may be more readily explained in galaxies where bipolar superwinds are present. Disc galaxies similar to the Speca class \citep{Hota2011}, which contain extended relic radio lobes, may provide particularly favourable conditions for such a mechanism. The source RAD J131346.9+500320 may represent a candidate example of this scenario. Embedded within this large-scale emission are two prominent ring-like structures with characteristic sizes of about 300 kpc. These rings appear to expand roughly along the major axis of the host galaxy, which is also aligned with a jet-like feature emerging from the central radio component.

\section{Multi-wavelength studies}
\subsection{Radio properties of circular diffuse sources}

Circular diffuse radio sources, including ORCs, exhibit morphologies that are distinct from those of classical radio galaxies or cluster relics. Many of these systems display circular or slightly elliptical structures with edge-brightened rims and relatively faint emission within their interiors. The radio emission is often concentrated along the outer boundaries, forming circular-symmetry morphologies. In contrast, the central regions generally lack prominent jet-like structures or bright radio cores, indicating that the currently observed emission may not be directly associated with ongoing jet activity.

The angular extent of such sources typically ranges from about one to two arcminute, corresponding to 100--500~kpc in projection \citep{No21b, No21c, Ko21, Om22c, Lo23, Ku23a, Ku23b, Ku25, koribalski2024, Ko24a, No25} except for RAD J131346.9+500320 (800 kpc). Their radio spectra generally show steep synchrotron emission with spectral indices in the range $\alpha \approx -0.8$ to $-1.5$ {(S$_{\nu}\sim\nu^{\alpha}$}), suggesting non-thermal radiation produced by relativistic electrons \citep{No21b, No21c, Ko21, Om22c, Lo23, Ku23a, Ku23b, Ku25, koribalski2024, Ko24a, No25}.

\begin{figure}
\centering
\includegraphics[width=0.8\textwidth]{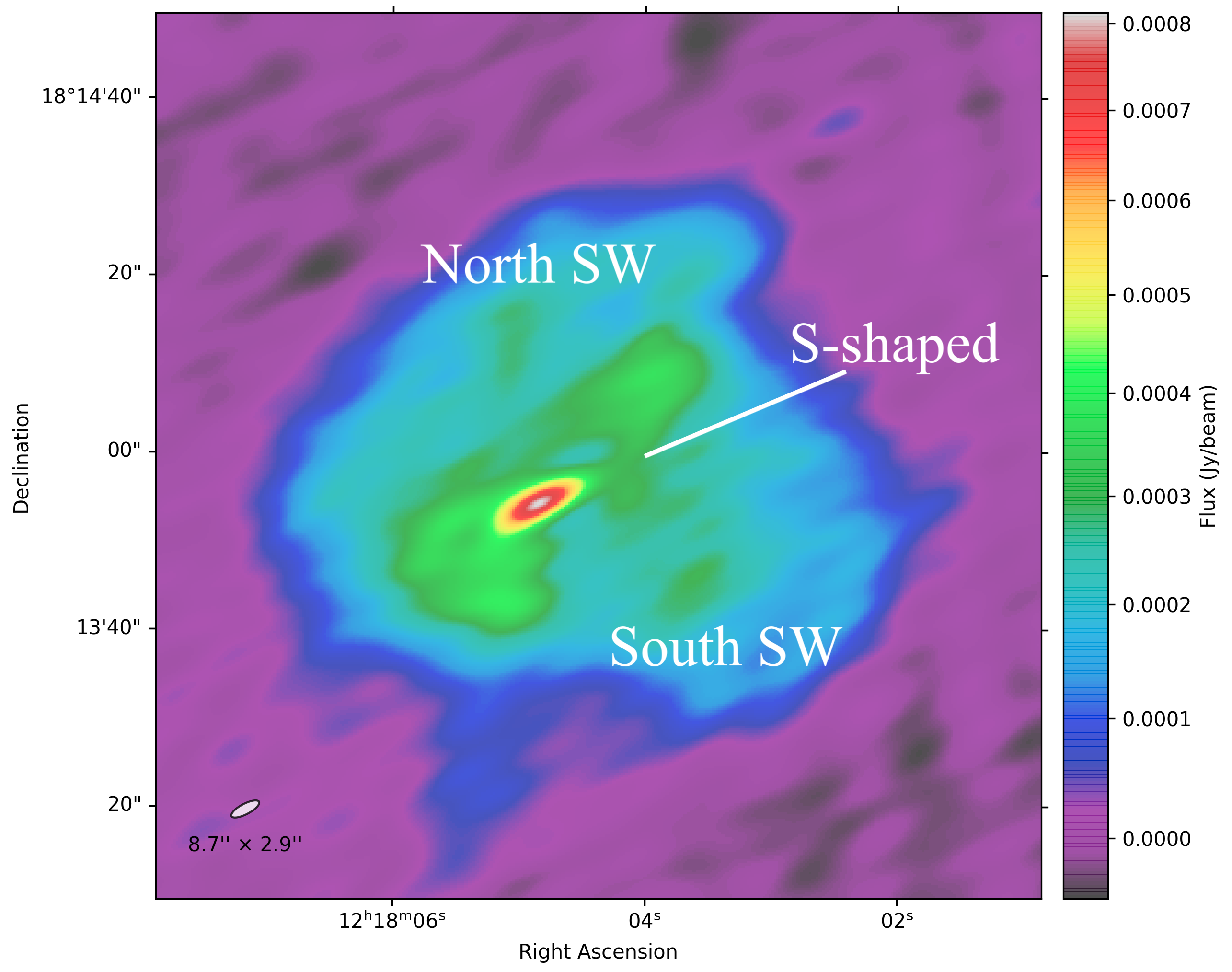}
\caption{VLA C-band (4–8 GHz) radio continuum image of the circular diffuse source J1218+1813. The high-resolution map reveals a resolved central core surrounded by faint diffuse emission with an S-shaped internal structure. Additional low-surface-brightness extensions are visible to the north and south (North SW and South SW), highlighting the complex internal morphology of circular diffuse radio sources adapted from \citet{Ku25}. The synthesized beam size is $8\farcs7 \times 2\farcs9$.}
\label{fig:VLA_C_J1218+1813}
\end{figure}

To investigate the detailed morphology and internal structure of circular diffuse radio sources, we carried out high-resolution radio observations of the circular diffuse sources J1407+0453, J1507+3013, and J1218+1813 using the VLA in the C-band (4--8 GHz) in the C configuration. Here we are presenting the high-resolution image of J1218+1813 in Figure~\ref{fig:VLA_C_J1218+1813} at 6 GHz. The final analysed image achieves an angular resolution of $8\farcs7 \times 2\farcs9$, enabling us to resolve the central radio core and trace the surrounding diffuse emission. The typical rms noise level of the map is approximately $11~\mu$Jy~beam$^{-1}$, estimated from off-source background regions, and the integrated flux density of the diffuse component is about 0.35 mJy at 6 GHz.

The diffuse radio emission extends over an angular diameter of nearly 70 arcsecond, corresponding to a projected physical size of $\sim 180$ kpc at the redshift $z = 0.139635$ of the host galaxy. The C-band VLA image reveals an inner patchy emission with an S-shaped morphology in J1218+1813. The southeast (SE) lobe shows a C-shaped bending, while the northwest (NW) side appears more irregular and diffuse.

\begin{figure}
    \centering
    \includegraphics[width=10cm]{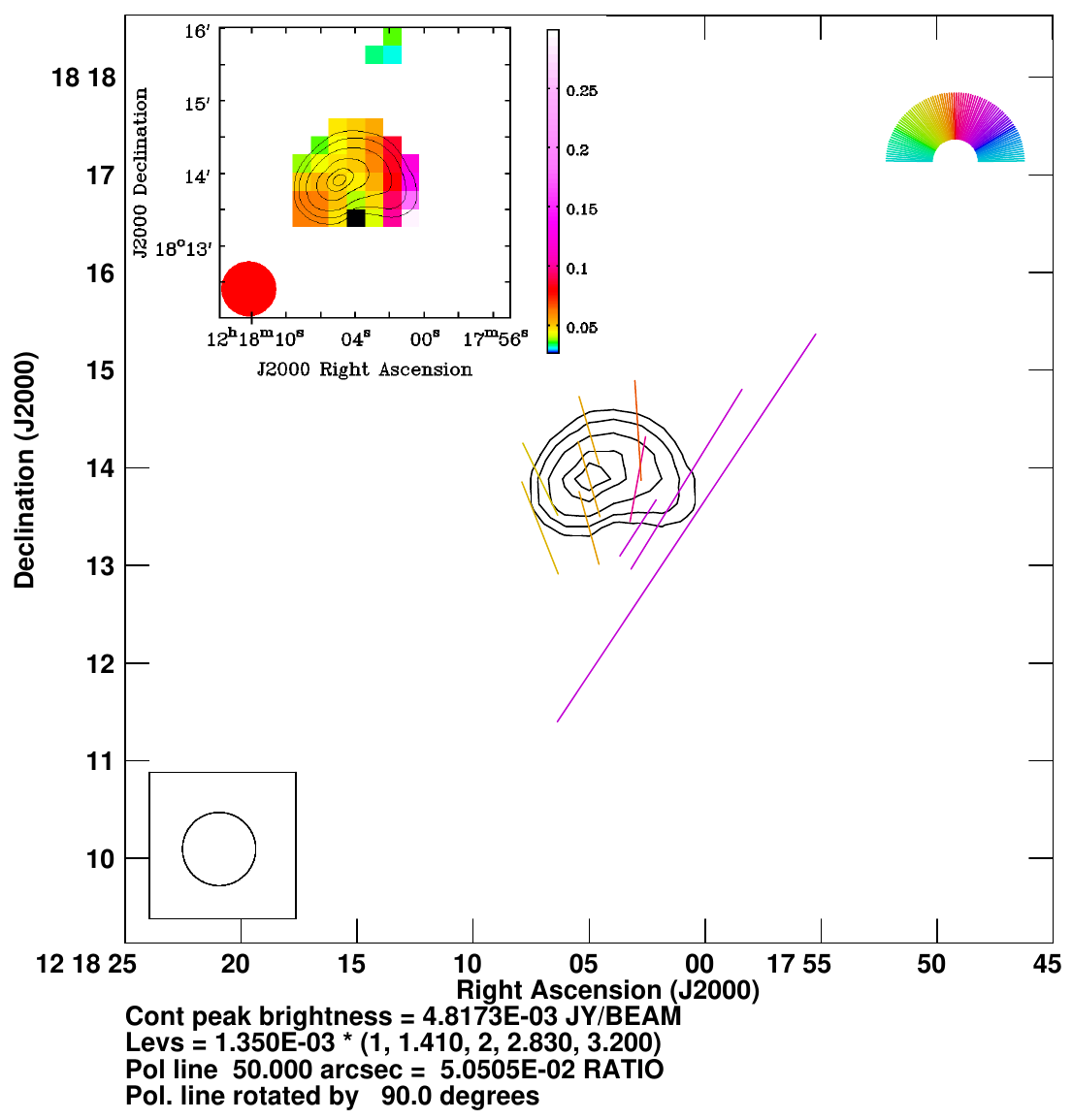}
    \caption{Polarization map of the circular diffuse radio source J1218+1813 adapted from \citet{Ku25}. Radio contours are overlaid with polarization vectors indicating the orientation of the projected magnetic field. Such polarimetric measurements provide important constraints on the magnetic field structure and particle acceleration processes within diffuse radio plasma.}
    \label{fig:polarization_J1218+1813}
\end{figure}

In addition to the main emission structure, faint wing-like extensions are visible on both sides of the source (labelled North SW and South SW in Figure \ref{fig:VLA_C_J1218+1813}), extending beyond the central emission region. These low-surface-brightness features indicate a complex internal radio structure within the diffuse circular emission.

\subsection{Polarization studies}
Observations of a few ORCs have revealed polarized emission along their rims \citep{2022MNRAS.513.1300N, Ta25}, indicating ordered magnetic fields associated with the radio-emitting plasma. For the diffuse circular radio source J1218+1813, a preliminary polarization map was produced using the National Radio Astronomy Observatory VLA Sky Survey \citep[NVSS;][]{Condon1998} (see Figure~\ref{fig:polarization_J1218+1813}). Polarization vectors are overlaid on the intensity contours to indicate the orientation of the projected polarization angles. The polarization vectors have been rotated by $90^\circ$ to represent the inferred magnetic field direction. The vector scale corresponds to $50''$ per polarization unit with a fractional polarization of $5\%$. Due to the limited sensitivity of the available data, the map provides only a preliminary characterization of the polarized emission associated with this diffuse circular radio source. Higher-sensitivity observations and wider frequency coverage will be necessary to obtain reliable polarization measurements and to better constrain the magnetic field properties of these faint, diffuse radio sources.

\subsection{Optical and infrared studies}
\begin{figure}
\centering
\includegraphics[width=0.7\textwidth]{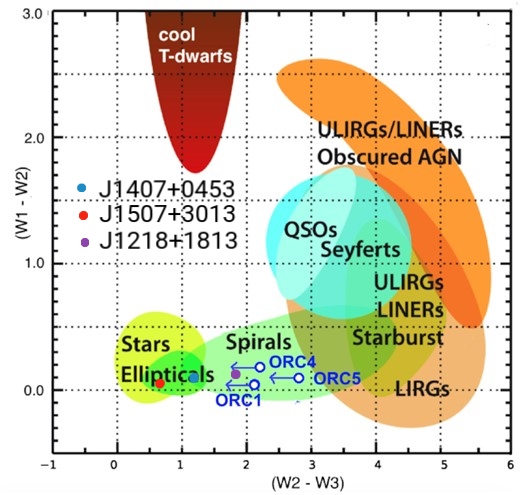}
\caption{WISE colour–colour diagram ($(W1-W2)$ versus $(W2-W3)$) showing the mid-infrared properties of host galaxies associated with circular diffuse radio sources. The diagram compares the positions of sources such as ASKAP-discovered ORCs (ORC1, ORC4, ORC5) and J1407+0453, J1507+3013, and J1218+1813 with typical galaxy populations. The positions of J1407+0453, J1507+3013, and J1218+1813 are marked on the standard WISE color–color diagram (adapted from \citet{2022MNRAS.513.1300N}).}
\label{fig:wise_color}
\end{figure}

Many confirmed ORCs are associated with faint optical galaxies, often consistent with massive early-type systems \citep[e.g.,][]{No21b, No21c, Ko21, koribalski2024, Ko24a, No25}. Infrared observations provide additional constraints on the nature of these hosts by tracing stellar populations and possible nuclear activity. The \textit{WISE} color--color diagram shown in Figure~\ref{fig:wise_color} compares the mid-infrared properties of several circular diffuse radio sources and ORCs with known galaxy populations. The diagram uses the \textit{WISE} $(W1-W2)$ versus $(W2-W3)$ color indices to classify galaxy types. The hosts of ORC1, ORC4, ORC5 (blue-white points) and J1218+1813 (purple filled point) lie within the region occupied by spiral galaxies. In contrast, the host galaxies of J1407+0453 and J1507+3013 (blue and red filled points) fall closer to regions populated by elliptical galaxies. This distribution suggests that circular diffuse radio sources may be associated with relatively passive elliptical galaxies or spiral galaxies.

\section{Formation scenarios}
Although the observational properties of circular diffuse radio sources and ORCs are becoming increasingly well documented, their physical origin remains uncertain. Several mechanisms have been proposed to explain the formation of these large-scale ring-like radio structures. The most commonly discussed scenarios involve energetic feedback from active galactic nuclei (AGN), large-scale galactic outflows, or shocks generated during galaxy group or cluster interactions \citep{No21b, No21c, Ko21, Om22c, Lo23, koribalski2024, Ko24a, No25}.

\subsection{AGN-driven shocks}
One of the most widely discussed formation scenarios involves large-scale shocks generated by AGN outflows. In this picture, powerful jets or winds launched from the central supermassive black hole inflate large plasma bubbles within the surrounding circumgalactic or intergalactic medium. As these bubbles expand, they can drive shock fronts into the ambient gas. Compression of magnetic fields and acceleration of relativistic electrons along these expanding fronts may produce synchrotron emission that appears as circular or shell-like radio structures.

If the AGN undergoes episodic activity, multiple outbursts could generate successive expanding shells, potentially giving rise to multiple ring-like features. Observationally, many circular diffuse radio sources exhibit edge-brightened rims and steep radio spectra with spectral indices typically in the range $\alpha \sim -1.0$ to $-1.5$, consistent with synchrotron emission from ageing relativistic plasma. Polarization measurements in several ORCs have also revealed magnetic field orientations aligned along the outer edges of the rings, which is consistent with magnetic field compression along expanding shock fronts.

Numerical simulations, such as those presented by \citet{Li24}, demonstrate that AGN-driven shocks propagating through the circumgalactic medium can generate diffuse circular radio structures with sizes of several hundred kpc. Such scales are comparable to those observed in known circular diffuse radio sources or ORCs. For systems that host galaxies showing evidence of past or current nuclear activity, the AGN-driven shock scenario therefore remains one of the most plausible explanations.

\subsection{Starburst-driven outflows}
An alternative explanation involves large-scale winds driven by intense star formation activity. In this scenario, stellar winds and supernova explosions collectively power galactic-scale outflows that expand into the surrounding circumgalactic medium \citep{Ve05, He90}. These outflows can inflate bubbles of hot gas and relativistic particles, analogous to the superbubbles observed in nearby starburst galaxies such as M82 \citep{Strickland2004}. Shock fronts associated with these winds can accelerate particles to relativistic energies, producing diffuse synchrotron emission detectable at radio wavelengths \citep{Lacki2013, Thompson2006}. If the starburst activity declines, the expanding bubble may continue to evolve as a low surface-brightness, circular radio structure.

However, several observational constraints challenge this scenario for many known ORCs and circular diffuse radio sources. Starburst-driven superbubbles typically extend over scales of a few tens of kpc, whereas circular radio structures detected in recent surveys often reach diameters of 200--700 kpc \citep{No21b, No21c, Ko21, Om22c, Lo23, Ku23a, Ku23b, koribalski2024, Ko24a, Ku25, No25}. Furthermore, many ORC host galaxies lack strong infrared emission indicative of intense star formation. As noted by \citet{coil2023}, these differences in scale and host-galaxy properties make the starburst-driven model less likely for the majority of currently known systems. Nevertheless, this mechanism may still operate in cases where circular radio emission is associated with galaxies undergoing significant star-forming activity.

\subsection{Galaxy group or cluster mergers}
Another possible formation mechanism involves shocks generated during interactions between galaxy groups or clusters. When large-scale structures merge, the resulting gravitational interactions can drive powerful shocks through the intragroup or intracluster medium. These shocks may re-accelerate pre-existing relativistic electrons and amplify magnetic fields, producing diffuse synchrotron emission.

In this context, circular radio structures could represent the projected appearance of merger-driven shock fronts propagating through the surrounding medium. Similar processes are known to produce radio relics in galaxy clusters, although typically on larger scales and with more elongated morphologies. Environmental asymmetries and density variations within the surrounding medium may lead to distortions or partial ring-like structures in the resulting radio emission.

Observations indicating that some circular diffuse radio sources reside in galaxy groups or moderately overdense environments lend some support to this scenario \citep{koribalski2024}. Detailed optical and spectroscopic studies of galaxies in the vicinity of these systems are required to determine whether large-scale dynamical interactions play a significant role in their formation. If confirmed, the merger-driven shock model would link circular diffuse radio sources to the broader context of structure formation and feedback processes in group-scale environments.

\begin{table*}
\scriptsize
\centering
\caption{Summary of proposed formation mechanisms for circular diffuse radio sources and their expected observational diagnostics. The table highlights morphological, spectral, polarization, and environmental signatures predicted by each scenario, together with observing strategies and feasibility under SKAO AA* and AA4 configurations.}

\begin{tabular}
{
>{\raggedright\arraybackslash}p{1.5cm}
>{\raggedright\arraybackslash}p{1.8cm}
>{\raggedright\arraybackslash}p{1.4cm}
>{\raggedright\arraybackslash}p{1.5cm}
>{\raggedright\arraybackslash}p{1.3cm}
>{\raggedright\arraybackslash}p{1.5cm}
>{\raggedright\arraybackslash}p{1.5cm}
>{\raggedright\arraybackslash}p{1.5cm}
>{\raggedright\arraybackslash}p{1.7cm}
}
\toprule
\textbf{Mechanism} &
\textbf{Morphology} &
\textbf{Spectral} &
\textbf{Pol./RM} &
\textbf{Host} &
%\textbf{Key Test} &
\textbf{Observations} &
\textbf{AA*/ AA4} &
\textbf{Targets} \\
\midrule

AGN shocks &
Circular shells, edge-brightened rims, possible concentric rings &
Steep ($\alpha\sim-1$), spectral ageing &
Ordered rim fields &
Massive galaxies, relic AGN &
%Shell thickness, spectral gradients &
Continuum + polarization &
AA*: detect; AA4: detailed &
Bright circular sources with AGN signatures \\
Starburst winds &
Bubble-like or spherical emission centered on star-forming galaxy &
Moderately steep spectra &
Weak/ turbulent fields &
IR-bright starbursts &
%SFR correlation &
Radio + IR + optical &
AA*: limited &
Low-$z$ star-forming galaxies with radio halos \\

Merger shocks &
Distorted or partial rings, asymmetric shells &
Steep, local flattening &
Patchy shock fields &
Groups/ clusters &
%Merger environment &
Radio + X-ray + optical &
AA*: partial; AA4: full &
Cluster/group merger systems \\

Fossil plasma (phoenix) &
Diffuse, filamentary or patchy emission, no clear shell &
Ultra-steep, spectral curvature &
Weak/ chaotic fields &
Old radio remnants &
%No active AGN &
Low-frequency radio surveys &
AA*: weak; AA4: strong &
Known relic lobes or fossil plasma regions \\

\bottomrule

\end{tabular}
\label{tab:mechanism}
\end{table*}

\section{Science enabled by the Square Kilometre Array Observatory}
The SKAO will bring a major improvement in our ability to study circular diffuse radio sources, including ORCs. It will be significantly more sensitive than current radio telescopes, with an improvement of about 4--10 times at low frequencies (50--350 MHz) and up to $\sim$ 10 times at higher frequencies (350 MHz--15 GHz). In addition, SKAO will provide a much larger field of view, increasing by a factor of $\sim$ 20 at frequencies between 350 MHz and 1.5 GHz, enabling faster and more efficient sky surveys. The angular resolution will also improve substantially, by up to a factor of $\sim$ 20 across the frequency range from 50 MHz to 15 GHz. SKA-MID will further extend observations to higher frequencies, up to $\sim$ 25 GHz or beyond.

These capabilities will allow SKAO to detect much fainter and more extended radio emission than is currently possible, and over much larger areas of the sky. As a result, it will enable systematic studies of circular diffuse radio sources across large cosmological volumes.

Even in its early AA* stage, SKAO will already be capable of discovering new sources. However, the full AA4 configuration will provide the sensitivity and angular resolution required for detailed studies, including morphological, spectral, and polarimetric characterization of these sources.

\subsection{Population studies of circular diffuse radio sources}
One of the most important contributions of SKAO will be the detection of a statistically significant population of circular diffuse radio sources. Current discoveries, based primarily on ASKAP and MeerKAT surveys, number only around ten objects \citep{No21b, No21c, Ko21, Om22c, Lo23, koribalski2024, Ko24a, No25, Ta25, 2025MNRAS5431048H}. The expected continuum sensitivity of $\sim$1~$\mu$Jy~beam$^{-1}$ for SKA-MID surveys will allow the detection of diffuse radio emission with surface brightness levels significantly below the detection thresholds of current surveys \citep{Braun2019}.

Given that typical circular diffuse sources have angular sizes of $\sim1-2$ arcminutes and steep radio spectra, SKAO surveys are expected to detect similar structures over a wide redshift range. This will enable the construction of the first large statistical samples of circular radio structures, allowing studies of their luminosity distribution, cosmological evolution, and environmental dependence.

\subsection{Morphological and structural studies}
The high angular resolution of SKA-MID will enable detailed imaging of the internal structures of circular diffuse radio sources. Sub-arcsecond imaging will allow the identification of morphological features such as partial shells, embedded inner rings, filamentary structures, or multiple concentric rings. While basic morphological classification can be performed using survey data, detailed structural studies will require deeper targeted observations with SKAO. The follow-up observations will provide important constraints on the dynamical processes responsible for shaping these structures. For example, resolving the thickness of radio rims or detecting internal radio features will help distinguish between expanding shell structures and diffuse relic plasma distributions.

\subsection{Spectral and polarization diagnostics}

The wide instantaneous bandwidth of SKAO will enable spatially resolved spectral index measurements across circular diffuse radio sources. Broadband imaging will allow the construction of spectral index maps, providing direct measurements of spatial gradients in synchrotron emission. In particular, spectral steepening can be used to trace radiative ageing of relativistic electrons, while localized deviations from smooth gradients may indicate variations in the underlying electron energy distribution.

Polarimetric observations provide a complementary probe of the magneto-ionic properties of diffuse radio plasma. Measurements of polazised flux density, fractional polarization, polarization position angles, and rotation measure (RM) will allow the characterization of magnetic field structure and Faraday effects across these extended objects. However, current observations of circular diffuse sources are often limited by low surface brightness and insufficient sensitivity, resulting in weak or marginal polarization detections \citep{2022MNRAS.513.1300N, Ku25, Ta25}.

The improved sensitivity, angular resolution, and frequency coverage of SKAO will enable high-fidelity polarization imaging and RM synthesis across these sources, allowing spatially resolved studies of magnetic field geometry and depolarization. Combined spectral and polarization diagnostics will therefore provide key constraints on the physical conditions, evolution, and interaction of diffuse radio plasma with the surrounding intergalactic medium. While wide-area surveys will enable the detection of such systems, detailed spectral and polarimetric analyses will generally require targeted follow-up observations.

\subsection{Science outcomes under AA* and AA4 configurations}
The staged deployment of SKAO will enable progressively deeper studies of circular diffuse radio sources. While the early AA* configuration will already allow the discovery of new systems and basic characterization of their morphology, the full AA4 baseline will provide the sensitivity, angular resolution, and polarimetric capability required for detailed physical investigations. Table~\ref{tab:AAcomparison} summarizes the key science outcomes achievable under the AA* and AA4 configurations.

\subsection{Quantitative discovery forecasts with SKAO}
Future wide-area radio surveys with the SKAO will significantly expand the discovery space for circular diffuse radio sources. Current ASKAP and MeerKAT surveys typically reach rms sensitivities of $\sim10$--$20\,\mu$Jy\,beam$^{-1}$ at angular resolutions of $\sim10''$--$15''$, which limits the detection of very low surface-brightness structures. In contrast, planned SKA-MID continuum surveys are expected to achieve rms sensitivities of $\sim1\,\mu$Jy\,beam$^{-1}$ at $\sim1''$--$2''$ resolution, depending on the survey strategy and integration time \citep{Braun2019}. 

For extended radio structures with angular sizes of $\sim1-2$ arcminutes, the detectable surface-brightness sensitivity scales with the synthesized beam size and the uv-coverage of the interferometer. Under typical SKA-MID survey conditions, diffuse emission with surface brightness levels of a few $\mu$Jy\,arcsec$^{-2}$ should become detectable, substantially improving the ability to identify faint circular radio structures that remain below the detection limits of current surveys.

Assuming that the presently known population represents only the brightest end of a much larger underlying distribution, order-of-magnitude extrapolations from existing ASKAP surveys suggest that SKA surveys could detect tens to hundreds of circular diffuse radio sources across large sky areas.

The wide instantaneous bandwidth of SKAO will also enable spatially resolved spectral index measurements across these structures. For sources with typical flux densities of a few hundred $\mu$Jy, broadband observations in SKA-MID spanning $0.35$--$15.4$ GHz can yield spectral index uncertainties of $\Delta\alpha \lesssim 0.1$ across resolved regions. In addition, the high polarization purity and sensitivity of SKAO will allow measurements of polarized emission at levels of a few percent of the total intensity, enabling detailed studies of magnetic field structures and Faraday rotation within these diffuse systems.

These capabilities will make it possible to distinguish between competing formation scenarios by combining morphological, spectral, and polarimetric diagnostics across a statistically significant population of circular diffuse radio sources.

\begin{table}
%\scriptsize
\centering
\caption{Comparison of science outcomes achievable with the staged AA* deployment and the full AA4 baseline configuration of SKAO.}
\begin{adjustbox}{width=1.0\textwidth}
\begin{tabular}{lll}

\hline
\textbf{Science Goal} & \textbf{AA* Capability} & \textbf{AA4 Capability} \\
\hline
Detecting circular sources 
& Detection of bright systems ($\gtrsim$ 10--50 $\mu$Jy beam$^{-1}$) 
& Detection of faint diffuse emission ($\sim$ 1 $\mu$Jy beam$^{-1}$) and large samples \\

Morphological classification 
& Identification of global structure ($\sim$ 10$''$ scale) 
& Detailed substructure ($\lesssim$ 1$''$ resolution), ring thickness, internal features \\

Spectral index mapping 
& Integrated or low-resolution spectra ($\Delta\alpha \sim 0.2$--0.3) 
& Spatially resolved spectral maps ($\Delta\alpha \lesssim 0.1$), ageing gradients \\

Polarization studies 
& Detection of polarized emission in bright regions ($\gtrsim$ 5--10\%) 
& High-fidelity polarization mapping, RM synthesis, magnetic field structure \\

Environmental studies 
& Nearby ($z \lesssim 0.5$) host identification 
& Cosmological population studies ($z \sim 1$--2), environment dependence \\

\hline

\end{tabular}
\end{adjustbox}
\label{tab:AAcomparison}
\end{table}
\section{Summary}

Circular diffuse radio sources, including ORCs, represent a recently identified class of large-scale synchrotron-emitting structures that challenge current models of radio galaxy evolution and feedback. Their circular morphologies, edge-brightened emission, and steep radio spectra distinguish them from classical radio galaxies and cluster-related diffuse sources, indicating that they trace distinct physical processes operating on scales of $\sim$ 200--500 kpc.

Multi-wavelength observations have revealed a diversity of structures, including single rings, inner-ring systems, diffuse circular emission with internal radio features, and double-ring morphologies. Radio, optical, infrared, and polarization data together provide initial constraints on their physical nature, suggesting a connection with large-scale energy injection and the evolution of relativistic plasma in low-density environments. However, the limited number of known sources and the lack of detailed spectral and polarimetric measurements currently prevent a definitive understanding of their origin.

This chapter has presented a systematic overview of circular diffuse radio sources and their observational properties, and has highlighted the key diagnostics required to distinguish between proposed formation scenarios. In particular, we emphasize the roles of morphology, spectral index variations, polarization properties, and host-galaxy environments as primary observational constraints.

The chapter is summarized as follows:

\begin{itemize}
\item Circular diffuse radio sources exhibit significant morphological diversity, including ring-like, inner-ring, and multi-ring structures, indicating that they are unlikely to form a homogeneous class.

\item Current observations suggest that these systems are associated with aged or re-energized synchrotron plasma, but the relative importance of different formation mechanisms (AGN-driven shocks, starburst outflows, merger-induced shocks, or revived fossil plasma) remains uncertain.

\item Spectral and polarization diagnostics provide key constraints on the evolution of relativistic electrons and the structure of magnetic fields, but existing data are limited by sensitivity and resolution.

\item The present sample of known sources is small, preventing statistically robust studies of their population properties and cosmological evolution.
\end{itemize}

Future observations with the SKAO will be transformative for this field. In particular, SKAO will enable:

\begin{itemize}
\item Detection of low surface-brightness circular diffuse radio sources down to $\sim \mu$Jy\,beam$^{-1}$ sensitivity levels, significantly expanding the known population.

\item High-resolution imaging ($\lesssim 1''$) to resolve internal substructures such as inner rings, filamentary features, and shell thickness.

\item Broadband spectral index mapping to measure spatial gradients and constrain radiative ageing and particle acceleration processes.

\item High-fidelity polarimetric imaging and RM synthesis to characterize magnetic field geometry and Faraday effects across diffuse radio plasma.

\item Systematic studies of host galaxy environments and redshift evolution, enabling the first statistically meaningful constraints on their formation and evolution.
\end{itemize}

In the AASKAII framework, the staged AA* configuration will enable the discovery of new circular diffuse radio sources and the identification of their basic morphological properties, while the full AA4 baseline will provide the sensitivity and resolution required for detailed spectral and polarimetric diagnostics. Together, these capabilities will allow direct testing of formation scenarios through observational constraints.

In summary, circular diffuse radio sources represent a key emerging population in extragalactic radio astronomy. With the advent of SKAO, these systems will become powerful probes of large-scale feedback, the evolution of relativistic plasma, and the role of magnetic fields in shaping the intergalactic medium.

\bibliographystyle{abbrvnat-maxbibnames4}
\bibliography{chapter}

\end{document}